\documentclass[article, prl, longbibliography, superscriptaddress, twocolumn]{revtex4-2}

\usepackage{amsmath,amssymb,amstext, amsthm}
\usepackage{bbm}
\usepackage{physics}
\usepackage{float}
\usepackage{braket}
\usepackage{graphicx}
\usepackage{color}
\usepackage{hyperref}

\def\bra#1{\langle#1|} \def\ket#1{|#1\rangle}
\def\braket#1#2{\langle#1|#2\rangle}

\def\proj#1{\ket{#1}\!\bra{#1}}
\def\vec#1{\mathbf{#1}}
\def\id{{\mathbb I}}

\def\q0{\underline{0}}

\def\H{{\cal H}}

\def\C{{\mathbb C}}
\def\id{{\mathbb I}}

\def\H{{\cal H}}

\def\R{\mathbb{R}}

\begin{document}

%\title{Entaglement generation in the Diósi-Penrose model of classical gravity.}

\title{The Di\'osi-Penrose model of classical gravity predicts gravitationally induced entanglement}

\author{David Trillo}
\email{david.trillo@cunef.edu}
\affiliation{Mathematics Department, CUNEF Universidad. Calle Almansa 101, 28040 Madrid, Spain}

\author{Miguel Navascués}
\email{miguel.navascues@oeaw.ac.at}
\affiliation{Institute for Quantum Optics and Quantum Information (IQOQI) Vienna\\ Austrian Academy of Sciences, Boltzmanngasse 3, Wien 1090, Austria}

\begin{abstract}

We show that the dynamics of the Diósi-Penrose (DP) model of classical gravity can entangle the mechanical degrees of freedom of two separate particles. For standard experiments of gravitationally induced entanglement (GIE), we find that entanglement can be generated iff the particles are separated by a distance smaller than some limiting value $d_c$, proportional to the only free parameter of the DP model. Greater distances can be achieved through new experimental configurations, where the initial wave-functions of the particles are allowed to spread perpendicularly to the separation axis. Although the DP dynamics asymptotically drives the system to a separable state, we observe that, for reasonable experimental parameters, GIE can survive for more than a day. Our results therefore imply that GIE \emph{detection} is not enough to validate quantum gravity. Experimental tests of GIE \emph{dynamics} have nonetheless the potential to falsify the DP model.

%We conclude that new regimes of the DP parameter can be tested in GIE experiments, and that the sole detection of GIE does not necessarily imply that gravity is quantum: a quantitative analysis of the entanglement produced is required just to tell apart DP from the Schrödinger equation with gravitational potential.

\end{abstract}

\maketitle

%\section{Introduction}

It is widely believed that two interacting quantum systems cannot become entangled if the interaction is mediated by a classical system (\cite{yi2022spatial, torovs2024relativistic, bose2023massive, MV24, marchese2024newton, chevalier2020witnessing} among others). Recent experimental proposals to test whether gravity is classical or quantum through table-top experiments \cite{bosetalk,BMM+17, MV17} rely on this basic assumption. In such ``Gravitationally induced entanglement'' (GIE) experiments \cite{MV24}, two massive particles are prepared each in a superposition of position states. The two particles are let to evolve, interacting with each other only through the gravitational field. The presence of entanglement after a finite time is then taken as proof that the gravitational field was not classical \cite{BMM+17, MV17, marchese2024newton, chevalier2020witnessing, bose2023massive, torovs2024relativistic, yi2022spatial}.

There have been refinements of the original GIE experiment proposal that study how to best measure the entanglement, and how to reduce the Casimir-Polder effect \cite{YSH+22, VMBM20}. Other works sought to generalize the original no-GIE argument to theories beyond quantum physics \cite{GGH22}. More recently, new experiments have been proposed which do not rely on the measurement of entanglement \cite{LPP24}; instead, they directly test whether the underlying interaction induces a channel achievable through Local Operations and Classical Communication (LOCC). The LOCC assumption used in both GIE experiments and the more recent \cite{GGH22,LPP24} has been criticized before \cite{MP23, GDG24, hall2021comment, marchese2024newton, magdalena}, but it nonetheless stands as the heart of all arguments in favor of the implication GIE$\Rightarrow$Quantum gravity. 

In this work, we consider a well known model of classical gravity, developed by Di\'osi \cite{Dio89} and later by Penrose \cite{Pen96} in $1989$ and $1996$. Originally derived from heuristic arguments, it was later shown to be equivalent to hybrid classical (gravity)-quantum (matter) dynamics \cite{Dio11}. We show that, in the Newtonian regime, the dynamics of the Di\'osi-Penrose (DP) model can entangle the mechanical degrees of freedom of two separate massive particles. Hence the DP model, the quintessential theory of classical gravity interacting with quantum matter, does not comply with the LOCC assumption found in \cite{BMM+17, MV17, GGH22}. Most importantly, our work proves that even the experimental observation of GIE is not enough to invalidate classical gravity.

Nonetheless, we find that, with regards to GIE, the DP model makes predictions qualitatively different from usual quantum mechanics. For instance, for the standard GIE experimental proposal, GIE is only possible if the particles are separated by a distance smaller than a limit $d_c\approx 0.8501\sigma$, where $\sigma$ is a free parameter of the DP model. In addition, whereas in the quantum case the system rapidly oscillates between zero and maximum entanglement, the DP dynamics predicts that the system will slowly increase its entanglement up to a limiting value and then back to zero, where it shall remain. Under the DP model, GIE is therefore expected to have a finite lifetime, although, for reasonable values of the experimental parameters, we find that it could last more than a day. Such special features of DP GIE can be used to either disprove the DP model, or at least to disprove a new regime of values of $\sigma$.

This paper is structured as follows: first, we review the experimental setup of GIE, including the argument invoked by the authors of \cite{BMM+17} and \cite{MV17} to regard GIE as a signature of non-classicality.
Next, we introduce the DP model and explain why the LOCC assumption of \cite{BMM+17,MV17,GGH22} does not apply. We will then use the DP theory to model the GIE experiment, characterizing the experimental conditions under which it is possible to observe GIE. We next run some computations with concrete experimental parameters to estimate how long and how strong the GIE effect predicted by the DP model is expected to be. Finally, we present our conclusions.

%\section{Gravitationally induced entanglement}
\vspace{10pt}
\noindent\emph{Gravitationally induced entanglement}
\vspace{10pt}

\begin{figure}
\centering
\includegraphics[scale=0.5]{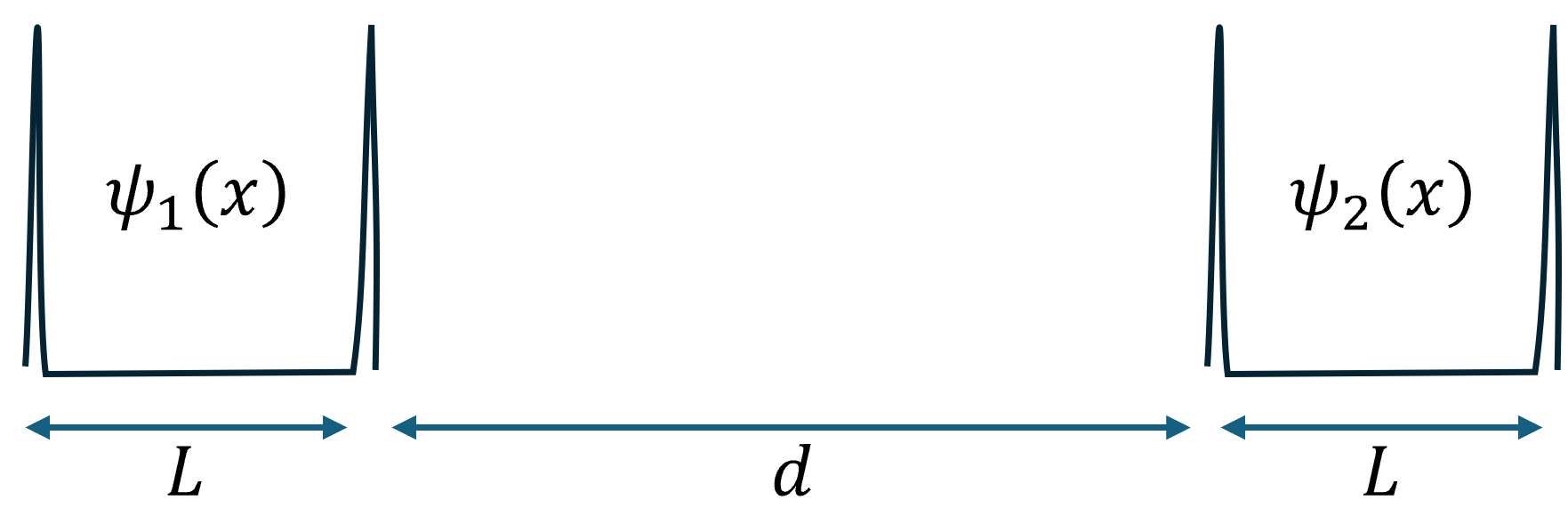}
\caption{Standard gravitationally induced entanglement experiment.}
\label{fig:horizontal}
\end{figure}

Past proposals to detect entanglement mediated by gravity consider experiments wherein two separate particles, initially prepared in a separable state, become entangled after some time. The starting position of the two particles, of masses $m_1$ and $m_2$, is assumed to be identical in two of the three spatial axes. As for the remaining one, say, the x axis, it is assumed that each particle is in a superposition of two different locations, see Figure \ref{fig:horizontal}. 

Let the initial states $\psi_1,\psi_2$ of particles $1,2$ be given by
\begin{equation}
\ket{\psi_1}:=\sum_{j=1}^2\alpha_j\ket{\vec{a}^j}, \ket{\psi_2}:=\sum_{j=1}^2\beta_j\ket{\vec{b}^j},
\label{initial_states}
\end{equation}
where $\vec{a}^j,\vec{b}^j\in\R^3$ are positions and, for any $\vec{c}\in\R^3$, $\ket{\vec{c}}$ denotes a wave-function centered at $x=\vec{c}$ with very low spread. The initial joint state of the system is therefore
\begin{equation}
\rho_0=\proj{\psi_1}\otimes\proj{\psi_2}.
\end{equation}
This is an instance of a \emph{separable state} \cite{entanglement_review}, i.e., a quantum state of the form 
\begin{equation}
\sum_ip_i\rho_i^1\otimes \rho_i^2,
\label{sep_state}
\end{equation}
where $p_i\geq 0$, for all $i$ and $\rho_i^1,\rho_i^2$ are normalized quantum states. Separable states are the only quantum states that can be prepared by two separate parties who can communicate through only a classical channel. In other words, they are realizable via local operations and classical communication (LOCC) \cite{entanglement_review}. %Indeed, any state of the form (\ref{sep_state}) can be LOCC prepared as follows: first, party $1$ randomly generates the symbol $i$ according to the distribution $\{p_i\}_i$. Next, she prepares state $\rho_i^1$ at her lab and uses the classical channel to ask party $2$ to prepare state $\rho_i^2$ on his side. If they both forget the value of $i$, then the prepared state is given by eq. (\ref{sep_state}). States that are not separable are said to be \emph{entangled} \cite{entanglement_review}.

Now, let $\vec{X}_i,\vec{P}_i$, respectively denote the vectors of position and momentum operators of particle $i \in \{1,2\}$. If we treat the gravitational interaction between the two particles the same way we model the Coulomb force, we would conclude that the two particles generate a gravitational field of the form:
\begin{equation}
\hat{\Phi}(\vec{r})=-G\int d\vec{x}\frac{\hat{\rho}(\vec{x})}{\|\vec{r}- \vec{x}\|},
\end{equation}
where $\hat{\rho}(\vec{x})=\sum_jm_j\delta(\vec{X}_j-\vec{x})$ is the mass density operator. For each value of $\vec{r}\in\R^3$, $\hat{\Phi}(\vec{r})$ is a non-trivial operator (namely, it is not proportional to the identity): we can therefore call it a \emph{quantum field}.

The two particles are thus subject to the potential $V_G$
\begin{equation}
\label{eq:std}
\hat{V}_G=\frac{1}{2}\int d\vec{r} \hat{\Phi}(\vec{r})\hat{\rho}(\vec{r})=-\frac{Gm_1m_2}{\|\vec{X}_1-\vec{X}_2\|}-\infty,
\end{equation}
which we can renormalize to eliminate the bothering (scalar) second term. In the absence of further forces, the two particles will hence jointly evolve according to the Hamiltonian $H_G=\sum_{i=1}^2\frac{\vec{P}_i^2}{2m_i}+\hat{V}_G$.

Call $\rho^G(t)=e^{-\frac{iH_Gt}{\hbar}}\rho(0)e^{\frac{iH_Gt}{\hbar}}$ the state of the system at time $t$. For sufficiently massive particles, we can neglect the contribution of the kinetic terms of $H_G$, in which case $\rho^G_{12}(t)=\proj{\psi(t)}$, with
\begin{equation}
\ket{\psi(t)}\approx\sum_{j,k=1}^2\alpha_j\beta_ke^{-i\frac{Gm_1m_2t}{\hbar\|\vec{a}^j-\vec{b}^k\|}}\ket{\vec{a}^j}\ket{\vec{b}^k}.
\end{equation}
The approximation will improve the narrower the initial wavepackets $\ket{\vec{a}^j}, \ket{\vec{b}^k}$ are. 

For $t>0$, the state $\rho_G(t)$ is, in general, entangled. To see this, we recall the following sufficient criterion for entanglement certification. 

Given any matrix $C\in B(\H_1\otimes \H_2)$, we define the linear map $C^{T_1}$ through the identity $(\ket{i}\bra{j}\otimes \ket{k}\bra{l})^{T_1}=\ket{j}\bra{i}\otimes \ket{k}\bra{l}$. Literally, we have transposed the first subsystem. The matrix $C^{T_1}$ is called the \emph{partial transpose} of $C$. As noted by Peres \cite{Peres96}, any separable state $\rho_{12}$ satisfies $\rho_{12}^{T_{1}}\geq 0$. Thus, if $\rho_{12}^{T_{1}}\not\geq 0$, then state $\rho_{12}$ is entangled. This method to detect entanglement is known as the \emph{Peres criterion} \cite{HORODECKI19961}.

Applying the Peres criterion to $\rho^G_{12}(t)$, the authors of \cite{BMM+17} find that the state becomes entangled if
\begin{equation}
\frac{Gm_1m_2t}{2\pi\hbar}\left(\frac{1}{d}+\frac{1}{d+2L}-\frac{2}{d+L}\right)\not\in\mathbb{Z}.
\end{equation}
That is, GIE will occur for almost all times $t$.

In \cite{BMM+17} and \cite{MV17} it is claimed that models of classical gravity do not predict GIE: any successful GIE experiment would thus constitute a refutation of all classical gravitational theories. The intuition is that, in experiments like the one just described, each massive particle would only interact with the gravitational field. If the latter happened to be classical, it could, at most, transfer classical information between the matter systems. Under the assumption of classical gravity, the quantum state of the particles is thus expected to undergo an LOCC operation, which cannot generate entanglement. As shown in \cite{GGH22}, this conclusion is sound even in scenarios where the matter systems are not described by quantum theory, but by a generalized probabilistic theory \cite{GPTs}.

There is, however, a flaw in the reasoning above: the dynamics of models of classical gravity do not necessarily decompose into sequences of pair-wise interactions between each massive particle and a classical field. In fact, as we show next, the most famous theory of classical gravity violates this fundamental assumption.

%\section{The Diósi-Penrose model}
\vspace{10pt}
\noindent\emph{The Diósi-Penrose model}
\vspace{10pt}

%The DP model is a collapse model \cite{BLS+13} where the collapse is due to a classical Newtonian gravitational interaction. As Tilloy has recently recontextualized the classification of classical-quantum interactions \cite{OSSW-D22} as systems with continuous measurement and feeback \cite{Tilloy24}, we can view the DP model as classical gravity mediating interaction between quantum systems \cite{TD16}.

The DP model is a theory of gravity, whereby matter is regarded as quantum and gravity, as a classical field. Both quantum and classical parts interact with each other in a non-trivial way. In the Newtonian approximation, the DP model can be derived as follows: let $\hat{\mu}_\sigma(\vec{x})d\vec{x}$ denote the Gaussian-smeared mass density at point $\vec{x}$, i.e.,
\begin{equation}
\label{eq:smeared}
\hat{\mu}_\sigma(\vec{x})\dd\vec{x}:=\frac{1}{(2\pi)^{3/2}\sigma^3}\sum_{j=1}^nm_je^{-\frac{\|\vec{x}-\vec{X}_j\|^2}{2\sigma^2}}\dd\vec{x},
\end{equation}
where the standard deviation $\sigma\in \R^+$ is a free parameter of the DP theory, introduced for renormalization purposes. 

We measure the operator $\hat{\mu}_\sigma(\vec{x})$ continuously at each point of space $\vec{x}$, call $\mu_t(\vec{x})=\langle \hat{\mu}_\sigma(\vec{x})\rangle_t +\delta \mu_\sigma(\vec{x})$ the (classical) outcome at time $t$. To model this process within the formalism of continuous measurements \cite{Wiseman_Milburn_2009}, we need to provide the state-independent correlations $\langle \delta \mu_t(\vec{x})\delta \mu_\tau(\vec{y})\rangle$, which take into account prior correlations between the measurement devices. The DP model takes those equal to $\delta(t-\tau)\frac{1}{8\pi G}\nabla^2\delta(\vec{x}-\vec{y})$ \cite{TD16}.

These outcomes are used to source a \emph{classical} field:
\begin{equation}
\Phi_c(\vec{r})=-G\int d\vec{x}\frac{\mu_t(\vec{x})}{\|\vec{x}-\vec{r}\|}.
\end{equation}
Integrating this field with the operator field $\frac{1}{2}\mu_\sigma(\vec{r})$ (in order to avoid the divergences appearing in (\ref{eq:std}) from construction), we arrive at a potential operator $\hat{W}_G$, which we use to back-react on the quantum particles. After tracing out the classical field, one arrives at the following master equation \cite{TD16}:
\begin{equation}
\label{eq:DP_evolution}
\frac{\dd\rho}{\dd t} = - \frac{i}{\hbar} [H,\rho] + \frac{G}{2 \hbar}A(\rho),
\end{equation}
where $H$ is the free Hamiltonian of the system, and 
\begin{align}
A(\rho)&=-\int \dd\vec{r} \dd\vec{s}\frac{1}{\|\vec{r}-\vec{s}\|}[\mu_\sigma(\vec{r}),[\mu_\sigma(\vec{s}),\rho]]\nonumber\\
&\qquad \quad+ i \int \dd\vec{r} \dd\vec{s} \frac{1}{\norm{\vec{r}-\vec{s}}} [\mu_\sigma(\vec{r})\mu_\sigma(\vec{s}),\rho],
\label{eq:diosi_hard}
\end{align}

Due to the structure of $\mu_\sigma(\vec{r})$, the operator $\hat{W}_G$ is local, i.e., it can be expressed as $\sum_j\hat{W}^j$, where $\hat{W}^j$ only acts non-trivially on system $j$. As such, it cannot generate entanglement by itself. However, the instruments carrying the prior weak measurement of $\hat{\mu}_\sigma(\vec{x})$ were assumed correlated. This measurement, together with the outcome-dependent back-reaction of $\hat{W}_G$, could in principle lead to entangled particle states, even if the particles participating in the experiment start in a separable state. That would not contradict the general no-go result of \cite{GGH22}, since the final DP dynamics (\ref{eq:DP_evolution}) might not factor into a sequence of pair-wise interactions between the gravitational field and one of the masses.

%\section{Occurrence of GIE in the DP model}
\vspace{10pt}
\noindent\emph{Occurrence of GIE in the DP model}
\vspace{10pt}

In this section, we study under which circumstances the two particles in Figure \ref{fig:horizontal} might become entangled by virtue of the Di\'osi-Penrose interaction (\ref{eq:DP_evolution}). 

First, in the Supplemental Material (section A), we prove that $A(\rho)$ admits a much simpler expression, namely:
\begin{equation}
\label{eq:DP_potential}
\bra{\vec{x}}A(\rho)\ket{\vec{y}} = g(\vec{x},\vec{y}) \bra{\vec{x}} \rho \ket{\vec{y}},
\end{equation}
where the function $g$ depends on the number of particles $n$, and their masses $\{m_j\}_j$ as
\begin{align}
g(\vec{x},\vec{y})&:= \sum_{j,k=1}^n m_j m_k \left[(i-1) f(\vec{x}_j,\vec{x}_k) \right.\nonumber\\
&\left. + (-i-1) f(\vec{y}_i,\vec{y}_k) + 2 f(\vec{x}_j,\vec{y}_k) \right].
\end{align}
Here $\vec{x}=(\vec{x}_1,...,\vec{x}_n)$, $\vec{y}=(\vec{y}_1,...,\vec{y}_n)$ denote the position vectors of the $n$ particles, and $f(\vec{z},\vec{z}'):=\tilde{f}(\|\vec{z}-\vec{z}'\|)$, with
\begin{equation}
\label{eq:ferf}
\tilde{f}(z) := \begin{cases} \erf\left( \frac{z}{2\sigma}\right)/z & \text{ if $z\not=0$} \\ \frac{1}{\sigma \sqrt{\pi}} & \text{ if $z=0$}, \end{cases}
\end{equation}
a function which depends on the parameter $\sigma$, and which can also be written as
\begin{equation}
\tilde{f}(z)= \frac{1}{\sigma \sqrt{\pi}} +  \frac{z^2}{12\sqrt{\pi}\sigma^3} + O(z^3/\sigma^4).
\end{equation}

The form (\ref{eq:DP_potential})-(\ref{eq:ferf}) is very convenient: if $H=0$, it implies that, for any set of $n$-particle coordinates $S\subset \R^{3n}$, an initial state $\rho_0$ with support in $\H=\overline{\mbox{span}}\{\ket{x}:x\in S\}$ will remain in $\H$ for all times.

Now, take $H=0$ in eq. (\ref{eq:DP_evolution}), and let $\rho_0=\proj{\bar{\psi}_1}\otimes \proj{\bar{\psi}_2}$,where $\bar{\psi}_1,\bar{\psi}_2$ denote the ``uniform'' wave functions
\begin{align}
&\ket{\bar{\psi}_1}:=\frac{1}{\sqrt{2}}(\ket{\vec{a}^1}+\ket{\vec{a}^2}),\nonumber\\
&\ket{\bar{\psi}_2}:=\frac{1}{\sqrt{2}}(\ket{\vec{b}^1}+\ket{\vec{b}^2}).
\label{uniform_states}
\end{align}
In accordance with Figure \ref{fig:horizontal}, we choose the initial positions to be $\vec{a}^1:=(0,0,0)$, $\vec{a}^2:=(L,0,0)$, $\vec{b}^1:=(d+L,0,0)$, $\vec{b}^2:=(d+2L,0,0)$. We also always assume from now on the masses of the two particles to be equal $m_1 = m_2 =: m$.

Given such an initial state, for $H=0$ each particle can be regarded as a qubit throughout the duration of the experiment. Thus, we use the standard quantum information notation 
\begin{equation}
\ket{0}_1:=\ket{\vec{a}^1}, \ket{1}_1:=\ket{\vec{a}^2},\ket{0}_2:=\ket{\vec{b}^1}, \ket{1}_2:=\ket{\vec{b}^2}.
\label{eq:qubit_reformulation}
\end{equation}
We also define the qubit pure states $\ket{\pm}:=\frac{1}{\sqrt{2}}(\ket{0}\pm\ket{1})$. 

Call $\rho(t)$ the solution of eq. (\ref{eq:DP_evolution}), with $H=0$ and $\rho(0)=\rho_0$. We next work out a sufficient criterion for the system to become entangled instantaneously, i.e., to ensure that $\rho(t)$ is entangled for $t\in(0,\epsilon)$ for some $\epsilon>0$.

Consider the Hermitian matrix
\begin{equation}
(\id-\proj{+}^{\otimes 2})\dot{\rho}(0)^{T_1}(\id-\proj{+}^{\otimes 2}).
\label{mato}
\end{equation}
Suppose that it has a negative eigenvalue, $\lambda$, with $\ket{\phi}$ being a corresponding eigenvector. Then, for small times $t$, $\rho(t)$ must be an entangled state.

Let us see why. Indeed, since $\ket{+}^{\otimes 2}$ is also an eigenvector of (\ref{mato}) with eigenvalue $0$, it must be orthogonal to $\ket{\phi}$. As $\ketbra{+}{+}^{\otimes 2}$ is invariant under partial transpositions, we have the identities
\begin{align}
&\bra{\phi}\rho(0)^{T_1}\ket{\phi}=|\langle \phi|+\rangle^{\otimes 2}|^2=0,\nonumber\\
&\frac{\dd}{\dd t}\bra{\phi}\rho(t)^{T_1}\ket{\phi}\Bigr|_{t=0}=\lambda,
\end{align}
It follows that $\bra{\phi}\rho(t)^{T_1}\ket{\phi}=\lambda t+O(t^2)$, and so $\rho(t)$ will be entangled for small $t$ by the Peres criterion.

Thanks to eq. (\ref{eq:DP_potential}), the eigenvalues of matrix (\ref{mato}) can be computed analytically. One of those is
\begin{align}
E_{-}:=&\frac{Gm^2}{2\hbar}\left(\frac{1}{\sqrt{\pi}\sigma}- \tilde{f}(L)\right.\nonumber\\
&\left.-\frac{1}{\sqrt{2}}\left|\tilde{f}(d+2L)+\tilde{f}(d)-2\tilde{f}(L+d)\right|\right).
\end{align}
Thus, if $E_{-}<0$, then $\rho(t)$ will be entangled for small values of $t$.

We verify that, indeed, $E_{-}$ can be negative for $d>0$. For instance, take $L=\frac{1}{8}\sigma, d=\frac{3}{8}\sigma$. Then we find $E_{-}=-0.0001915\frac{Gm^2}{2 \hbar\sigma}$. This proves that the Di\'osi-Penrose dynamics can be entangling. 

Furthermore, if $E_{-}<0$, the joint state will become entangled even if $H\not=0$, as long as $H$ is a local Hamiltonian (e.g.: a kinetic term). Indeed, call $\rho_H(t)$ the state evolved through eq. (\ref{eq:DP_evolution}) with local $H$ (with boundary conditions $\rho_H(0)=\rho(0)$), and consider the state in the interaction picture $\tilde{\rho}(t):=e^{iHt/\hbar}\rho_H(t)e^{-iHt/\hbar}$. Since $H$ is local, we have that $\tilde{\rho}(t)$ is entangled iff $\rho_H(t)$ is entangled. Now,
\begin{align}
\frac{\dd}{\dd t}\tilde{\rho}(t)&=\frac{i}{\hbar}[H,\tilde{\rho}(t)]-\frac{i}{\hbar}[H,\tilde{\rho}(t)]+\frac{G}{2\hbar}e^{iHt}A(\rho_H(t))e^{-iHt}\nonumber\\
&=\frac{G}{2\hbar}e^{iHt}A(\rho_H(t))e^{-iHt}.
\end{align}
Thus, we have that $\tilde{\rho}(0)=\rho(0)$, $\dot{\tilde{\rho}}(0)=\dot{\rho}(0)$. Hence, $\tilde{\rho}(t)^{T_1}$ has an eigenvalue of the form $E_{-}t+O(t^2)$. For small times $t$ it will therefore become entangled, and so will $\rho_H(t)$ by the argument above.

Unfortunately, for fixed $L$ there exists a distance $d_c(L)$ beyond which $E_{-}>0$. Numerically, we observe that $d_c(L)$ increases with decreasing $L$. Assuming that this is true, one should not expect to observe GIE for $d>d_c:=\lim_{L\to 0}d_c(L)$.

We next compute $d_c$. We observe that
\begin{equation}
\lim_{L\to 0}\frac{E_{-}}{L^2}=\frac{Gm^2}{2\hbar}\left(-\frac{1}{2}\ddot{\tilde{f}}(0)-\frac{1}{\sqrt{2}}|\ddot{\tilde{f}}(d)|\right)=:\gamma(d).
\end{equation}
$d_c$ is therefore the solution of the equation $\gamma(d_c)=0$, which gives $d_c\approx 0.850872\sigma$.

One could think that, even if $E_{-}\geq 0$, it might still be the case that the DP dynamics (\ref{eq:DP_evolution}) can entangle (perhaps other) separable quantum states with support in $\{\ket{\vec{a}^j}\ket{\vec{b}^k}\}_{j,k}$ (and perhaps not instantaneously). This is actually \emph{not} the case: as shown in the Supplemental Material (section B), for $H=0$ the DP dynamics is entangling iff $E_{-}<0$.

\begin{figure}
\centering
\includegraphics[scale=0.5]{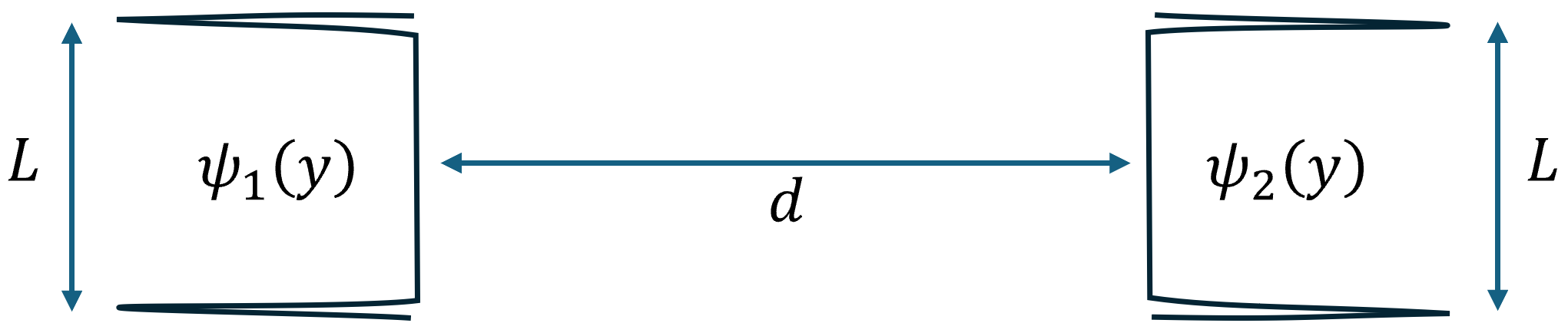}
\caption{Entanglement experiment, transversal configuration.}
\label{fig:vertical}
\end{figure}

Is there a way to observe GIE beyond $d_c$? Consider an alternative spatial configuration in which both initial wave-functions are spread in the $\hat{y}$ axis instead of the $\hat{x}$ axis, see Figure \ref{fig:vertical}. Namely: the first particle has support in $\{(0,0,0),(0,L,0)\}$, while the second particle has support in $\{(d,0,0),(d,L,0)\}$. Thus, any measurement of the particles' positions would reveal them to be at a distance at least $d$ from each other.

In the Supplemental Material (section C), we analyze this `transversal' scenario. We verify that GIE is observable iff 
\begin{equation}
\frac{1}{\sqrt{\pi}\sigma}-\tilde{f}(L)- \sqrt{2}(\tilde{f}(d)-\tilde{f}(\sqrt{L^2+d^2}))<0.    
\end{equation}
This time, we numerically observe that the greatest distances that allow GIE are achieved for very large $L$. In the limit $L\to \infty$,  we find that entanglement generation is possible if $d< \bar{d}_c\approx 2.21093\sigma$. The transversal configuration can therefore exhibit GIE at distances almost three times greater than $d_c$.

%\section{Quality of GIE in the DP model}
\vspace{10pt}
\noindent\emph{Quality of GIE in the DP model}
\vspace{10pt}

In the previous section, we proved that the Di\'osi-Penrose dynamics is entangling and characterized under which circumstances we shall expect to observe GIE. In this section, we study how much entanglement can be generated and for how long. For the sake of comparison with the standard proposals, we restrict ourselves to the configuration of Figure \ref{fig:horizontal}. A similar analysis can be made in the transversal scenario without much difficulty.

We will assume that $H=0$, since in that case the equations of motion are easily integrated. This a usual approximation done in the analysis of the GIE experiments, and so the regime of validity is similar. Everything will work the same if $H\propto P$, but studying the case $H\propto P^2$ would require very involved numerical integration. 

Under this assumption, the equations of motion (\ref{eq:DP_evolution}), (\ref{eq:DP_potential})-(\ref{eq:ferf}) are easily integrated to be
\begin{equation}
    \bra{\vec{x}}\rho(t)\ket{\vec{y}}:= e^{\frac{G}{2\hbar} g(\vec{x},\vec{y})t} \bra{\vec{x}} \rho(0) \ket{\vec{y}}.
\end{equation}
That is, in general the state decoheres exponentially to a mixture of position eigenvectors. This, however, can lead to entanglement generation during some time. Indeed, by choosing $\rho(0)$ as in the previous section, we may suppose that we are in a Hilbert space of dimension two (eqs. (\ref{uniform_states}) and (\ref{eq:qubit_reformulation})) by approximating all other matrix elements as 0, as is usually done in the analysis in GIE experiments. As in the previous section, to certify entanglement we verify the existence of a negative eigenvalue in the partial transpose of $\rho(t)$ (the "negativity" \cite{negativity}), which is easily computed from all the equations. 

\begin{figure}
\centering
\includegraphics[clip, trim=2cm 12.5cm 5cm 0.5cm, scale=0.6]{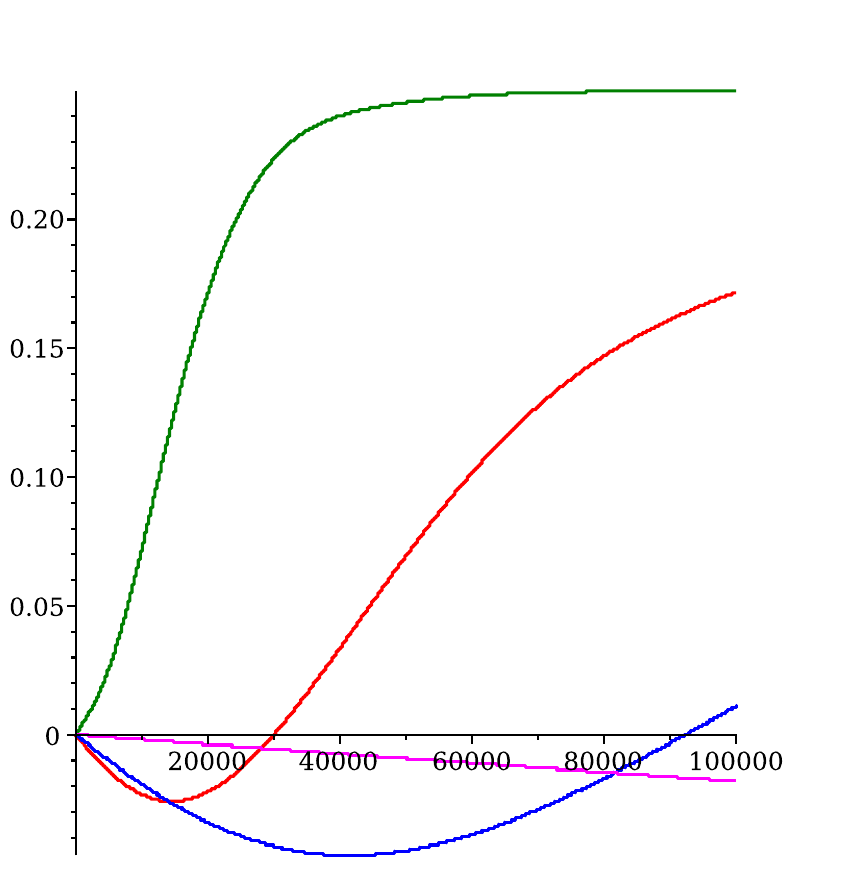}
\caption{Smallest eigenvalue of the transpose of the state (negativity) vs time elapsed, for four different choices of the DP parameter  [$\sigma_{\text{Green}} = 50\mu m, \sigma_{\text{Red}}=75\mu m, \sigma_{\text{Blue}}=100 \mu m, \sigma_{\text{Magenta}}=250 \mu m$], with $L= 23 \mu m, d=24\mu m, m_1=m_2=10^{-15}kg$, time in seconds (length and mass parameters chosen from the experimental proposal \cite{VMBM20}). With these parameters, the negativity predicted by linearized gravity, with the same initial conditions would have oscillated between $0$ and $-0.5$ around $200$ times.}
\label{fig:ent_finite}
\end{figure}

Figure \ref{fig:ent_finite} shows a plot of the smallest eigenvalue of $\rho(t)^{T_1}$ in SI units.
Of note is the long sustained entanglement that appears, which is related to the poor ability of the DP model to decohere fast enough, as discussed in \cite{FDG-R+24}. Of course, eventually all the generated entanglement will be destroyed as the state becomes maximally decohered. The parameters (except $\sigma$) were chosen from the proposal \cite{VMBM20}, which shows that indeed with the proper choice of the free parameter in the DP model, we can get entanglement generation in proposed GIE experiments. Nevertheless, the kind of entanglement obtained has a very different signature. DP GIE looks like an overdamped version of the entanglement predicted by linearized gravity. When measured with a witness, we find that the preferred witness for the usual GIE proposals \cite{CPK20} does not actually detect the DP GIE, at least in the horizontal configuration and this parameter range.

We also observe that, with these length and mass parameters, noticeable entanglement is generated only for a small, but previously untested range of values of $\sigma$. A GIE experiment is, therefore, useful for ruling out (or validating) the DP model. By changing the length and mass scales, and spatial configuration in the experiment, it is possible to explore a variety of regimes for $\sigma$. 

%\david{So, now that I wrote this and I cleaned my code a little, I quickly checked with the original parameters of ref[1], for which I got no entanglement originally with $10^{-4}$. Indeed, you only get entanglement for sigma from $7.5\cdot10^{-4}$ on, and it's like $\sim -0.05$ again. So it seems that the maximum entanglement is this at most? The entanglement builds up faster and lasts for less time at least. I've left this in the "cleancode" in github. So the worse the parameters the larger the sigmas you are testing. Maybe mention this somehow?}

%\begin{figure}[H]
%\centering
%\includegraphics[scale=0.5]{lmao.png}
%\caption{Smallest eigenvalue of the transpose vs time elapsed, for $L= 23 \mu m, d=24\mu m, m_1=m_2=10^{-15}kg, \sigma = 10^{-4}m$, time in seconds. DP vs Schrödinger comparison}
%\label{fig:ent_finite}
%\end{figure}

%\section{Conclusion}
\vspace{10pt}
\noindent\emph{Discussion}
\vspace{10pt}

We have proven that the dynamics of the DP model of classical gravity can generate entanglement between two separate massive systems, provided that the parameters of the experimental setup and the free constant $\sigma$ of the DP model are chosen appropriately. Therefore, experimental proposals to test the quantumness of gravity through GIE need to be analyzed with care. Until now, the DP model was only seen as a possible source of decoherence and noise that would make it harder to detect entanglement \cite{BGU17}, but it turns out that it can be a source of entanglement as well. So, it could very well be that if entanglement is detected in a GIE experiment, the contribution comes from spontaneous collapse due to a classical gravitational field, rather than from a quantum aspect of gravity. As we observed, the magnitude and the dynamics of GIE predicted by the DP model and the usual Schrödinger equation differ considerably. This means that, if the world is better approximated by the latter, then future experiments of GIE might falsify the DP model.

If that were the case, could we then conclude that gravity is a quantum field? Not necessarily. Even if the DP model cannot fit GIE experimental data, there is a large set of classical-quantum candidate models that could \cite{OSSW-D22, Tilloy24}. Future research aiming to disprove classical gravity should tackle them all.

\vspace{10pt}
\noindent\emph{Code availability}
\vspace{10pt}

All the computer codes used in this manuscript can be found in \cite{codigo}.

%\section*{Acknowledgements}
\vspace{10pt}
\noindent\emph{Acknowledgements}
\vspace{10pt}

M.N. acknowledges useful discussions with Antoine Tilloy and Esteban Castro-Ruiz.

\bibliography{bib_re}

\newpage

%\section{Appendix}

\begin{appendix}

\section{Supplemental Material}
\section{Appendix A: Derivation of eqs. (13)-(15) in the main text}
\label{app:diosi_simple}

In this section, we are going to show how to go from equations (9)-(11) to equations (13)-(15) in the main text. This is a straightforward computation. The integrals we need to compute are all of the form
\begin{equation}
\int \dd\vec{r} \dd\vec{s} \frac{1}{\norm{\vec{r}-\vec{s}}} \mu_\sigma(\vec{r}) \mu_\sigma(\vec{s}) \rho,    \label{basic_int}
\end{equation}
for different permutations of the operators $\mu_\sigma(\vec{r})$,  $\mu_\sigma(\vec{s})$, $\rho$.
In position basis, we obtain for the value of $\bra{\vec{x}}\text{(eq. (\ref{basic_int}))}\ket{\vec{y}}$ the expression
\begin{align}
&\sum_{i,j=1}^n \frac{m_im_j}{(2\pi)^3\sigma^6} \int \dd\vec{r}\dd\vec{s} \frac{1}{\norm{\vec{r}-\vec{s}}}e^{-\frac{\norm{\vec{s}-\vec{x}_i}^2+\norm{\vec{r}-\vec{x}_j}^2}{2\sigma^2}}\bra{\vec{x}} \rho \ket{\vec{y}}    
\end{align}
and analogous ones for the other four terms in the expansion of eq. (10) in the main text. The problem thus reduces to computing the integral
\begin{equation}
I:= \int \dd\vec{r}\dd\vec{s} \frac{1}{\norm{\vec{r}-\vec{s}}}e^{-\norm{\vec{s}-\vec{x}}^2/2\sigma^2} e^{-\norm{\vec{r}-\vec{y}}^2/2\sigma^2}
\end{equation}
for some fixed vectors $\vec{x},\vec{y}$. The appropriate way to do this is via the change of variables
\begin{equation}
\begin{pmatrix} \vec{u} \\ \vec{v} \end{pmatrix} = \begin{pmatrix} 1 & -1 \\ 1 & 1\end{pmatrix} \begin{pmatrix} \vec{r} \\ \vec{s} \end{pmatrix}
\end{equation}
for which we obtain
\begin{equation}
I =\frac{e^{-(\vec{x}^2+ \vec{y}^2)/2\sigma^2}}{2^3} J_1 J_2,
\end{equation}
where
\begin{equation}
J_1:=\int \dd\vec{v} e^{-\vec{v}^2/4\sigma^2} e^{\vec{v}\cdot (\vec{x}+\vec{y})/2\sigma^2}
\end{equation}
and
\begin{equation}
J_2:= \int \dd\vec{u} \frac{1}{\norm{\vec{u}}} e^{-\vec{u}^2/4\sigma^2}e^{\vec{u}\cdot(\vec{x}-\vec{y})/2\sigma^2}.
\end{equation}
The first integral, $J_1$, is Gaussian and therefore easily evaluated after completing the square to
\begin{equation}
J_1 = (\sqrt{4\pi \sigma^2})^3 e^{(\vec{x}+\vec{y})^2/4\sigma^2}.
\end{equation}
For the second integral, we change to spherical coordinates so that $\dd\vec{u} = \lambda ^2 \sin(\theta)\dd\lambda \dd\theta \dd\varphi$. Without loss of generality, we may assume that $(\vec{x}-\vec{y})$ lies in the $z$-axis, so that
\begin{align*}
J_2 =& 2\pi \int_0^\infty \dd\lambda \lambda e^{-\lambda^2/4\sigma^2} \int_{0}^{\pi} \dd\theta\sin(\theta) e^{\lambda  \norm{\vec{x}-\vec{y}}\cos(\theta)/2\sigma^2} \\
=& 8\pi \sigma^2 \int_0^\infty \dd\lambda \frac{1}{\norm{\vec{x}-\vec{y}}} e^{-\lambda^2/4\sigma^2}\sinh(\lambda \norm{\vec{x}-\vec{y}}/2\sigma^2) \\
=& \frac{8 \sigma^3\sqrt{\pi}^3}{\norm{\vec{x}-\vec{y}}} e^{\norm{\vec{x}-\vec{y}}^2/4\sigma^2}\erf(\norm{\vec{x}-\vec{y}}/2\sigma).
\end{align*}
Therefore,
\begin{equation}
I=  \frac{(2\pi \sigma^2)^3}{\norm{\vec{x}-\vec{y}}}\erf(\norm{\vec{x}-\vec{y}}/2\sigma)
\end{equation}
The result follows from substituting this back in the original equations.

\section{Appendix B: Necessary and sufficient conditions for GIE in the standard scenario}
\label{app:charact_hori}
Consider two particles of identical mass $m$. At time $t=0$, we set them up in some separable state $\rho_0$ and make them evolve through eq. (9) in the main text with $H=0$. Call $\rho(t)$ the state of this two-particle system after time $t$ and suppose that, at time $t^-_{ent}$, the bipartite state becomes entangled. That is, $\rho(t)$ is separable for $t\in [0,t^-_{ent}]$ and entangled for $t\in (t^-_{ent},t^+_{ent})$, where $t^+_{ent}$ might be infinite. In that case, at $t=t_{ent}$, one of the pure product states in the separable decomposition of $\rho(t_{ent})$ must become entangled. We have just proven that, if any time-independent master equation is entangling, then there exists a pure product state $\rho_0=\proj{\psi_1}\otimes \proj{\psi_2}$, with $\psi_1,\psi_2$ of the form eq. (1) in the main text, such that the evolved state $\rho(t)$ becomes entangled for $t\in(0,\epsilon)$, for some $\epsilon>0$. For the time being, for any pair of normalized pure states $\ket{\phi_1},\ket{\phi_2}$ we denote by $\rho(t;\phi_1,\phi_2)$ the state of the system at time $t$, given that it was prepared at time $t=0$ in state $\proj{\phi_1}\otimes \proj{\phi_2}$.

Note that the Diosi-Penrose Lindbladian, eq. (13) in the main text, commutes with local operations of the form 
\begin{equation}
\rho\to (r(\vec{X}_1)\otimes s(\vec{X}_2))\rho (r(\vec{X}_1)\otimes s(\vec{X}_2))^\dagger,
\end{equation}
where $\vec{X}_1, \vec{X}_2$ respectively denote the vector of position operators of particles $1$ and $2$. Define $C_1=\alpha(\vec{X}_1)$ ($C_2=\beta(\vec{X}_2)$), where $\alpha:\R^3\to\C$ ($\beta:\R^3\to\C$) is any function such that $\alpha(\vec{a}^k)=\sqrt{2}\alpha_k$ ($\beta(\vec{b}^k)=\sqrt{2}\beta_k$) for $k=1,2$, where $\alpha_k,\beta_k$ are the coefficients of the states $\psi_1,\psi_2$, see eq. (1) in the main text. Then we have that
\begin{equation}
\rho(t;\psi_1,\psi_2)=(C_1\otimes C_2)\rho(t;\bar{\psi}_1,\bar{\psi}_2)(C_1\otimes C_2)^\dagger,
\label{local_op}
\end{equation}
where $\bar{\psi}_1,\bar{\psi}_2$ denote the ``uniform'' wave functions appearing in eq. (17) in the main text.

Since the transformation on the right-hand side of (\ref{local_op}) is local, it follows that, if $\rho(t;\ket{\psi_1},\ket{\psi_2})$ is entangled, then so is $\rho(t;\ket{\bar{\psi}_1},\ket{\bar{\psi}_2})$. Thus, if the Di\'osi-Penrose dynamics is entangling, then it must be able to entangle uniform states instantaneously. In order to study the potential for entanglement generation of the Diosi-Penrose dynamics, we can therefore choose the initial state of the joint system to be $\rho_0=\proj{\bar{\psi}_1}\otimes \proj{\bar{\psi}_2}$. From simplicity, from now we use $\rho(t)$ again to mean the solution of the equation of motion (\ref{eq:DP_evolution}), with initial conditions $\rho_0=\proj{\bar{\psi}_1}\otimes \proj{\bar{\psi}_2}$.

As it turns out, for two-qubit systems, the Peres criterion \cite{Peres96} is necessary and sufficient for separability \cite{HORODECKI19961}. Thus, all we need to verify is whether $\rho(t)^{T_1}\geq 0$ for small $t$. To this aim, we next compute the eigenvectors of $\rho(t)^{T_1}$ at different orders in perturbation theory. Call $\varphi_0(t),...,\varphi_3(t)$ the eigenvectors of the $4\times 4$ matrix $\rho^{T_1}(t)$, with eigenvalues $\lambda_0(t),...,\lambda_3(t)$. As it is usual in perturbation theory, we normalize the eigenstates so that
\begin{equation}
\braket{\varphi_i(0)}{\varphi_i(t)}=1, i=0,...,3,
\end{equation}
for all times $t$.

Expanding the equation
\begin{equation}
(\rho^{T_1}(t)-\lambda_i(t))\ket{\varphi_i(t)}=0,    
\end{equation}
on powers of $t$, we obtain the identities
\begin{align}
&(\rho^{T_1}(0)-\lambda_i(0))\ket{\varphi_i(0)}=0,\nonumber\\
&(\rho^{T_1}(0)-\lambda_i(0))\ket{\dot{\varphi}_i(0)}+(\dot{\rho}^{T_1}(0)-\dot{\lambda}_i(0))\ket{\varphi_i(0)}=0,\nonumber\\
&\frac{1}{2}(\rho^{T_1}(0)-\lambda_i(0))\ket{\ddot{\varphi}_i(0)}+(\dot{\rho}^{T_1}(0)-\dot{\lambda}_i(0))\ket{\dot{\varphi}_i(0)}\nonumber\\
&+\frac{1}{2}(\ddot{\rho}^{T_1}(0)-\ddot{\lambda}_i(0))\ket{\varphi_i(0)}=0.
\label{pert_rels}
\end{align}
The first identity of eq. (\ref{pert_rels}) implies that $\{\ket{\varphi_i(0)}\}_i$ are eigenvectors of $\rho^{T_1}(0)=\proj{+}^{\otimes 2}$. This matrix obviously has a positive eigenvalue $\lambda_0(0)=1$, with eigenvector $\ket{\varphi_0(0)}=\ket{+}^{\otimes 2}$; the remaining eigenvalues $\lambda_1(0),\lambda_2(0),\lambda_3(0)$ are zero. Thus, the zeroth order conditions are not enough to fix $\ket{\varphi_1(0)},\ket{\varphi_2(0)},\ket{\varphi_3(0)}$.

Now, set $i\in\{1,2,3\}$ and act on the second line of (\ref{pert_rels}) with $\bra{\varphi_j(0)}$, with $j=1,2,3$. We arrive at:
\begin{equation}
\bra{\varphi_j(0)}\dot{\rho}^{T_1}(0)\ket{\varphi_i(0)}=\dot{\lambda}_i(0)\delta_{ij}.
\end{equation}
Define $P:=\sum_{i=1}^3\proj{\varphi_i(0)}=\id-\proj{+}^{\otimes 2}$. Then, the equation above is equivalent to
\begin{equation}
P\dot{\rho}^{T_1}(0)P\ket{\varphi_i(0)}=\dot{\lambda}_i(0)\ket{\varphi_i(0)}.
\end{equation}
Thus, $\dot{\lambda}_1(0),\dot{\lambda}_2(0),\dot{\lambda}_3(0)$ are the eigenvalues of matrix (19) (in the main text). Leaving aside the $\ket{+}^{\otimes 2}$ subspace, we find that this matrix has one $0$ eigenvalue, with eigenvector $\ket{-}^{\otimes 2}$, and the two eigenvalues
\begin{align}
E_{\pm}:=&\frac{Gm^2}{2\hbar}\left(\frac{1}{\sqrt{\pi}\sigma}- \tilde{f}(L)\right.\nonumber\\
&\left.\pm\frac{1}{\sqrt{2}}\left|\tilde{f}(d+2L)+\tilde{f}(d)-2\tilde{f}(L+d)\right|\right).
\end{align}
Let us (arbitrarily) assign $\dot{\lambda}_1(0)=E_{+}$, $\dot{\lambda}_2(0)=E_{-}$, $\dot{\lambda}_3(0)=0$, in which case $\ket{\varphi_3(0)}=\ket{-}^{\otimes 2}$.

We set $i=3$ and act on the third line of eq. (\ref{pert_rels}) with $\bra{\varphi_3(0)}$. The result is:
\begin{align}
&\frac{1}{2}\left(\bra{\varphi_3(0)}\ddot{\rho}^{T_1}(0)\ket{\varphi_3(0)}-\ddot{\lambda}_3(0)\right)\nonumber\\
&+\bra{\varphi_3(0)}\dot{\rho}^{T_1}(0)\ket{\dot{\varphi}_3(0)}=0.
\label{lambda_3dotdot}
\end{align}
To evaluate the last term, we take $i=3$ in the second line of eq. (\ref{pert_rels}) and act on the ket with $\bra{\varphi_0(0)}$, obtaining:
\begin{equation}
\braket{\varphi_0(0)}{\dot{\varphi}_3(0)}+\bra{\varphi_0(0)}\dot{\rho}(0)^{T_1}\ket{\varphi_3(0)}=0.
\label{overlap_conn}
\end{equation}
Therefore, we have that
\begin{align}
&\bra{\varphi_3(0)}\dot{\rho}^{T_1}(0)\ket{\dot{\varphi}_3(0)}\nonumber\\
&=\bra{\varphi_3(0)}\dot{\rho}^{T_1}(0)P\ket{\dot{\varphi}_3(0)}\nonumber\\
&+\bra{\varphi_3(0)}\dot{\rho}^{T_1}(0)\ket{\varphi_0(0)}\braket{\varphi_0(0)}{\dot{\varphi}_3(0)}\nonumber\\
&=-|\bra{\varphi_3(0)}\dot{\rho}^{T_1}(0)\ket{\varphi_0(0)}|^2,
\end{align}
where we invoked the identity $P+\proj{\varphi(0)}=\id$ in the first equality and relations $\bra{\varphi_3(0)}\dot{\rho}(0)^{T_1}P=0$ and (\ref{overlap_conn}) in the second.

Thus, by eq. (\ref{lambda_3dotdot}), we have that
\begin{align}
&\ddot{\lambda}_3(0)=\bra{-}^{\otimes 2}\ddot{\rho}^{T_1}(0)\ket{-}^{\otimes 2}\nonumber\\
&-2|\bra{-}^{\otimes 2}\dot{\rho}^{T_1}(0)\ket{+}^{\otimes 2}|^2=:2\nu.
\end{align}
We verify that
\begin{align}
\nu=&\frac{G^2m^4}{8\hbar^2}\left[2\left(\frac{1}{\sqrt{\pi}\sigma}-\tilde{f}(L)\right)^2\right.\nonumber\\
&\left.+\left(\tilde{f}(d)+\tilde{f}(2L+d)-2\tilde{f}(L+d)\right)^2\right].
\end{align}

In summary, the eigenvalues of $\rho(t)^{T_1}$ are of the form $\{1+O(t),E_{\pm}t+O(t^2),\nu t^2+O(t^3)\}$, with $\nu(d,L)$ non-negative and non-zero for $L>0$.

All the above proves that entanglement can be generated if $E_{-}<0$, and it cannot be generated if $E_{-}>0$. Now, suppose that $E_{-}=0$. Then, the state at some small time $t$ might in principle be entangled, yet through an arbitrarily small variation of $d,L$ we could make $E_{-}>0$, in which case the state at time $t$ will be separable. Since $\rho(t)$ is continuous on $d,L$, that would imply that there exist separable states arbitrarily close to $\rho(t)$, which, by the closure of the set of separable states, implies that $\rho(t)$ itself is separable. Therefore, for $H=0$, the experiment in Figure 1 in the main text can exhibit GIE iff $E_{-}<0$.

\section{Appendix C: Sufficient conditions for GIE in the transversal scenario}
\label{app:ent_trans}
To determine if the Diosi-Penrose dynamics can generate entanglement in this scenario, it suffices to assume that the initial state of the two particles is
\begin{equation}
\frac{1}{2}\left(\ket{(0,0,0)}+\ket{(0,L,0)}\right)\otimes \left(\ket{(d,0,0)}+\ket{(d,L,0)}\right)
\end{equation}

In that case, the eigenvalues of $\rho(t)^{T_1}$ are verified to be of the form $1+O(t)$ (with eigenvector $\ket{+}\otimes\ket{+}$), $\bar{\nu}t^2+O(t^3)$ (with eigenvector $\ket{-}\otimes\ket{-}$), and $\bar{E}_{\pm}t+O(t^2)$, with
\begin{align}
&\bar{\nu}=\frac{G^2m^4}{4\hbar^2}\left[\left(\frac{1}{\sqrt{\pi}\sigma}-\tilde{f}(L)\right)^2+2\left(f(d)-\tilde{f}(\sqrt{a^2+L^2})\right)^2\right],\nonumber\\
&\bar{E}_{\pm}:=\frac{Gm^2}{2\hbar}\left(\frac{1}{\sqrt{\pi}\sigma}-\tilde{f}(L)\pm \sqrt{2}(\tilde{f}(d)-\tilde{f}(\sqrt{L^2+d^2}))\right).
\end{align}
As before, $\bar{\nu}$ turns out to be a non-negative, non-vanishing analytic function of $d,L$. Thus, the arguments from Section \ref{app:diosi_simple} apply and, for $H=0$, the DP dynamics is entangling iff $\bar{E}_-(d,L)<0$.

However, in this case we numerically observe that $\bar{E}_-(d,L)<0$ for some $L$ iff $\lim_{L\to\infty}\bar{E}_-(d,L)<0$. Computing the last quantity, we arrive at the expression:
\begin{equation}
\frac{Gm^2}{2\hbar}\left(\frac{1}{\sqrt{\pi}\sigma} -\sqrt{2} \tilde{f}(d)\right).
\end{equation}
Equalling it to $0$, we find that entanglement generation is possible as long as $d\leq \bar{d}_c\approx 2.21093\sigma$.
\end{appendix}

\end{document}